\documentclass[12pt]{article}
\usepackage{amsfonts}
\usepackage{amssymb}
\def\hybrid{\topmargin -20pt    \oddsidemargin 0pt
        \headheight 0pt \headsep 0pt
        \textwidth 6.25in       % A4 paper
        \textheight 9.25in       % A4 paper
        \marginparwidth .875in
        \parskip 5pt plus 1pt   \jot = 1.5ex}

\hybrid

\def\baselinestretch{1.2}

\catcode`\@=11

\def\marginnote#1{}
\newcount\hour
\newcount\minute
\newtoks\amorpm
\hour=\time\divide\hour by60
\minute=\time{\multiply\hour by60 \global\advance\minute by-\hour}
\edef\standardtime{{\ifnum\hour<12 \global\amorpm={am}%
        \else\global\amorpm={pm}\advance\hour by-12 \fi
        \ifnum\hour=0 \hour=12 \fi
        \number\hour:\ifnum\minute<10 0\fi\number\minute\the\amorpm}}
\edef\militarytime{\number\hour:\ifnum\minute<10 0\fi\number\minute}
\def\draftlabel#1{{\@bsphack\if@filesw {\let\thepage\relax
   \xdef\@gtempa{\write\@auxout{\string
      \newlabel{#1}{{\@currentlabel}{\thepage}}}}}\@gtempa
   \if@nobreak \ifvmode\nobreak\fi\fi\fi\@esphack}
        \gdef\@eqnlabel{#1}}
\def\@eqnlabel{}
\def\@vacuum{}
\def\draftmarginnote#1{\marginpar{\raggedright\scriptsize\tt#1}}

\def\draft{\oddsidemargin -.5truein
        \def\@oddfoot{\sl preliminary draft \hfil
        \rm\thepage\hfil\sl\today\quad\militarytime}
        \let\@evenfoot\@oddfoot \overfullrule 3pt
        \let\label=\draftlabel
        \let\marginnote=\draftmarginnote
   \def\@eqnnum{(\theequation)\rlap{\kern\marginparsep\tt\@eqnlabel}%
\global\let\@eqnlabel\@vacuum}  }

\def\preprint{\twocolumn\sloppy\flushbottom\parindent 2em
        \leftmargini 2em\leftmarginv .5em\leftmarginvi .5em
        \oddsidemargin -.5in    \evensidemargin -.5in
        \columnsep .4in \footheight 0pt
        \textwidth 10.in        \topmargin  -.4in
        \headheight 12pt \topskip .4in
        \textheight 6.9in \footskip 0pt
        \def\@oddhead{\thepage\hfil\addtocounter{page}{1}\thepage}
        \let\@evenhead\@oddhead \def\@oddfoot{} \def\@evenfoot{} }

\def\numberbysection{\@addtoreset{equation}{section}
        \def\theequation{\thesection.\arabic{equation}}}

\def\underline#1{\relax\ifmmode\@@underline#1\else
        $\@@underline{\hbox{#1}}$\relax\fi}

\def\titlepage{\@restonecolfalse\if@twocolumn\@restonecoltrue\onecolumn
     \else \newpage \fi \thispagestyle{empty}\c@page\z@
        \def\thefootnote{\fnsymbol{footnote}} }

\def\endtitlepage{\if@restonecol\twocolumn \else \newpage \fi
        \def\thefootnote{\arabic{footnote}}
        \setcounter{footnote}{0}}  %\c@footnote\z@ }

\catcode`@=12
\relax

\def\figcap{\section*{Figure Captions\markboth
        {FIGURECAPTIONS}{FIGURECAPTIONS}}\list
        {Figure \arabic{enumi}:\hfill}{\settowidth\labelwidth{Figure
999:}
        \leftmargin\labelwidth
        \advance\leftmargin\labelsep\usecounter{enumi}}}
 \relax
\def\tablecap{\section*{Table Captions\markboth
        {TABLECAPTIONS}{TABLECAPTIONS}}\list
        {Table \arabic{enumi}:\hfill}{\settowidth\labelwidth{Table
999:}
        \leftmargin\labelwidth
        \advance\leftmargin\labelsep\usecounter{enumi}}}
 \relax
\def\reflist{\section*{References\markboth
        {REFLIST}{REFLIST}}\list
        {[\arabic{enumi}]\hfill}{\settowidth\labelwidth{[999]}
        \leftmargin\labelwidth
        \advance\leftmargin\labelsep\usecounter{enumi}}}
 \relax
\makeatletter
\newcounter{pubctr}
\def\publist{\@ifnextchar[{\@publist}{\@@publist}}
\def\@publist[#1]{\list
        {[\arabic{pubctr}]\hfill}{\settowidth\labelwidth{[999]}
        \leftmargin\labelwidth
        \advance\leftmargin\labelsep
        \@nmbrlisttrue\def\@listctr{pubctr}
        \setcounter{pubctr}{#1}\addtocounter{pubctr}{-1}}}
\def\@@publist{\list
        {[\arabic{pubctr}]\hfill}{\settowidth\labelwidth{[999]}
        \leftmargin\labelwidth
        \advance\leftmargin\labelsep
        \@nmbrlisttrue\def\@listctr{pubctr}}}
 \relax
\makeatother
\newskip\humongous \humongous=0pt plus 1000pt minus 1000pt

\newif\ifdtup

\relax

\def\be{\begin{equation}}
\def\ee{\end{equation}}
\def\ba{\begin{eqnarray}}
\def\ea{\end{eqnarray}}

\def\del{\partial}

\def\Tr{\mathrm{Tr}}

\def\a{\alpha}

\def\b{\beta}

\def\g{\gamma}

\def\d{\delta}
\def\D{\Delta}
\def\e{\epsilon}

\def\th{\theta}

\def\m{\mu}
\def\n{\nu}

\def\l{\lambda}

\def\s{\sigma}

\def\cN{{\cal N}}

\def\no{\noindent}

\def\qq{\qquad}

\def\IR{\relax{\rm I\kern-.18em R}}

\def \ha {{1\over 2}}

\def \ov {\over}

\def\II{\relax{\rm 1\kern-.35em1}}
\def\IR{\relax{\rm I\kern-.18em R}}
\def\inv{^{\raise.15ex\hbox{${\scriptscriptstyle -}$}\kern-.05em 1}}

\def\ii{\mathrm{i}}
\begin{document}
\renewcommand{\theequation}{\arabic{equation}}
\renewcommand{\theequation}{\thesection.\arabic{equation}}

\newcommand{\beq}{\begin{equation}}
\newcommand{\eeq}[1]{\label{#1}\end{equation}}
\newcommand{\ber}{\begin{eqnarray}}
\newcommand{\eer}[1]{\label{#1}\end{eqnarray}}
\newcommand{\eqn}[1]{(\ref{#1})}
\begin{titlepage}
\begin{center}

\hfill CCNY-HEP-10/4\\
%\hfill arXiv:yymm.nnnn [hep-th]\\
\vskip  0.5in

{\large \bf Solving field equations in non-isometric coset CFT backgrounds}
%\\ and the power of hidden symmetry}
%or

%{\large \bf Putting non-isometric coset CFT backgrounds to work}

%or

\vskip 0.4in

{\bf Alexios P. Polychronakos$^1$}\phantom{x} and\phantom{x}
{\bf Konstadinos Sfetsos}$^{2}$ \vskip 0.1in

${}^1\!$ Physics Department, City College of the CUNY\\
160 Convent Avenue, New York, NY 10031, USA\\
{\footnotesize{\tt alexios@sci.ccny.cuny.edu}}

\vskip .2in

${}^2\!$
Department of Engineering Sciences, University of Patras\\
26110 Patras, GREECE\\
{\footnotesize{\tt sfetsos@upatras.gr}}\\

\end{center}

\vskip .3in

\centerline{\bf Synopsis}

\no
The largest known class of gravitational backgrounds with an exact string theoretical
description is based on coset $G/H$ CFTs and the corresponding gauged WZW models.
These backgrounds generically lack isometries and are quite complicated. Thus
the corresponding field equations seem impossible to solve and
their use in physical applications becomes problematic.
We develop a systematic general method enabling us to overcome this problem using
group theory.
The method is inspired by observations made in some elementary geometric coset and
coset CFTs, but its full power is apparent in non-abelian cases. We analyze exhaustively the coset
$SU(2) \times SU(2)/SU(2)$ and explicitly solve the
scalar wave equation of the corresponding gravitational background. We also examine the
high spin limit and derive the effective geometry
that consistently captures the corresponding sector in the theory.

%\vskip .4in
%\noindent
%August 2002\\
\vfil
% \begin{flushleft}
% NEIP--05--01 \\
% LPTENS--05/01\\
% CPTH-PC001.0105\\
% ROMA--1400/05\\
% January 2005
% \end{flushleft}

\end{titlepage}
\vfill
\newpage
\setcounter{footnote}{0}
\renewcommand{\thefootnote}{\arabic{footnote}}

\renewcommand{\theequation}{\thesection.\arabic{equation}}

%\setlength{\baselineskip}{.7cm} \setlength{\parskip}{.2cm}

%% Section 1:
\setcounter{section}{0}

\def\baselinestretch{1.2}
\baselineskip 20 pt %17.5 pt
\noindent

%%%%%%%%%%%%%%%%%%%%%%%%%%%%%%%%%%%%%%%%%

\tableofcontents

\section{Introduction}

Understanding and probing gravitational effects at the quantum level
beyond the General Theory of Relativity is an important issue.
String theory has contributed decisively in this by providing exact
models with a clear spacetime interpretation. The most appealing
class of such models is based on coset conformal field theory (CFT),
that is, models on a $G/H$ coset manifold \cite{coset} that admit
a spacetime interpretation via the gauged WZW models \cite{gwzwac}.
The extraction of the gravitational background fields follows a
well established procedure, initiated with the prototype example of
a two-dimensional black hole in \cite{WittenBH}.

\no
One major hurdle that has hindered the widespread usage of
three- and
higher-dimensional such models is their complexity, and, in
particular, the generic lack of isometries of
the corresponding gravitational backgrounds. In physical applications
one often has to deal with
field equations in order to extract physical information, and
the absense of isometries makes the exact solution of these
equations an impossible task with any of the traditional methods.
In the present paper we overcome this problem by using
methods based on the rich, albeit not manifest, underlying group theoretic structure.

\no
The present paper is organized as follows: In section 2 we review
relevant aspects of WZW and gauged WZW models. In section 3 we motivate
the idea with some elementary examples involving the $SU(2)$ group manifold and the
geometric and CFT cosets constructed by using a $U(1)$ subgroup.
We then present the general method for obtaining the solution
of the scalar wave equation for the general CFT coset
$G/H$ model.
In section 4 we test our general result by working out explicitly the details for the CFT
corresponding to the non-abelian $SU(2)\times SU(2)/SU(2)$ coset model.
We also derive the gravitational background that
consistently describes the high spin sector of the theory.
Finally, in section 5 we present our conclusions as well as directions for future
related work.

\section{Geometrical aspects of gauged WZW models}
\setcounter{footnote}{0}
\renewcommand{\thefootnote}{\arabic{footnote}}

In this section we briefly review relevant aspects of WZW and gauged WZW models.

\subsection{Aspects of WZW models}

Consider a group $G$ and the associated Lie-algebra generators $\{T_A\}$, $A=1,2,\dots , \dim(G)$, obeying
the commutation rules
\be
[T_A,T_B]= i f_{AB}{}^C T_C\ ,
\ee
with structure constants $f_{AB}{}^C$.
The WZW action for a group element $g$ in some representation of $G$ is given by
\cite{WZ,Witten}
\be
S_{\rm WZW}(g) = k I_{0}(g)\ ,
\label{e-2-1}
\ee
where
\be
I_0(g) = {1\ov 2\pi} \int_M \Tr(\del_+ g^{-1} \del_- g) + {i \ov 6\pi} \int_M \Tr(g^{-1} dg)^3\ .
\label{e-2-2}
\ee
Here $\s^\pm$ denote light-cone coordinates on the worldsheet $M$ and $B$ is
a three-dimensional ball bounded by $M$.
The relative coefficient of the cubic term is completely dictated by the Polyakov--Wiegman identity \cite{PW}
\be
I_0(g_1 g_2) = I_0(g_1) + I_0(g_2) - {1\ov \pi} \int \Tr(g_1^{-1} \del_- g_1 \del_+ g_2 g_2^{-1})\ .
\label{powi}
\ee
One realizes an infinite dimensional current algebra symmetry \cite{Witten}
%\be
%g(\s^+,\s^-) \to \Omega_+ (\s^+) g(\s^+,\s^-) \Omega_- (\s^-)\ ,
%\label{e-2-3}
%\ee
generated by the (on-shell) chiral and anti-chiral currents
$J_+ = \del_+ g g^{-1}$ and $J_- = g^{-1} \del_- g$. The central extensions of this algebra is an
integer (for compact groups) $k$ that appears as an overall coefficient in the action \eqn{e-2-1}. (The quantization of $k$ can also be
realized as a topological quantization of the coefficient of the WZ
term in the action.)

\no
It will be useful to introduce the left- and right-invariant Maurer--Cartan forms with components
\be
L^A_M = - \ii \Tr (g^{-1} \partial_M g T^A)\ , \qq R^A_M = - \ii \Tr ( \partial_M g g^{-1} T^A)\ ,
\label{e-2-7}
\ee
as well as the obvious notation for the currents $L^A_\pm =  L^A_M \del_\pm X^M$ and
$R^A_\pm =  R^A_M \del_\pm X^M$, with the $X^M$'s being the parameters of the group element $g$ acting as
coordinates in the target space.
This procedure recasts the original action \eqn{e-2-1} into the form
\be
S_0(g) = {k \ov \pi} \int_M d^2 \s (G_{MN} + B_{MN}) \partial_- X^\m \partial_+ X^\n \ ,
\label{e-2-9}
\ee
where $G_{MN}$ is the $\s$-model metric, given for WZW models by
\be
G_{MN} = G^{(0)}_{MN} \equiv \ha \eta_{AB} L^A_M L^B_N = \ha \eta_{AB} R^A_M R^B_N\ ,
\label{e-2-10}
\ee
and $B_{MN}$ are the components of a two-form whose strength, defined by $H=d B$, reads
\be
H_{MNP}=H^{(0)}_{MNP} \equiv \ha f_{ABC} L^A_M L^B_N L^C_P = - \ha f_{ABC} R^A_M R^B_N R^C_P\ .
\label{e-2-11}
\ee
Note in passing that, even if an explicit parametrization of the element $g$ in terms of coordinates $X^M$ is given,
it is sometimes more convenient to bypass the computation of the left- and right-invariant
forms and directly compute the $\s$-model action by using repeatedly the Polyakov--Wiegman identity \eqn{powi}.

\subsection{Aspects of gauged WZW models}

\no
The construction of conformal coset backgrounds based on the gauged WZW models starts by
introducing gauge fields $A_\pm$ in the Lie-algebra of a subgroup $H\in G$.
These are used to gauge a subgroup of the symmetry group $G_L \times G_R$.
Restricting for simplicity to the diagonal
case, where $H$ is embedded in $G_L$ and $G_R$ in the same manner, one introduces the gauged WZW action
\cite{gwzwac}
\be
S_{\rm gWZW}(g,A_\pm) = k I_{0}(g)
+ {k\ov \pi} \int_M \Tr (A_- \del_+ g g^{-1} - A_+ g^{-1} \del_- g + A_- g A_+ g^{-1} - A_- A_+)\ .
\label{gwwzw}
\ee
This is invariant under the gauge transformations
\be
g\to h^{-1} g h \ ,\qq A_\pm \to h^{-1} A_\pm h - h^{-1}\del_\pm h\ ,\quad  {\rm for}\ h(\s^+,\s^-)\in H\ .
\label{ejhg11}
\ee

\no
The procedure of obtaining a $\s$-model involves two steps. Due to the gauge invariance we may gauge fix
$\dim(H)$ parameters in $g$, thus reducing the number of parameters to $\dim(G/H)$,
thereafter denoted by $X^\m$. Effectively, these are the
parameters that are gauge invariant \cite{BaSfe3} under the subgroup action \eqn{ejhg11}.
It is convenient to split the group indices as $A=(a,\a)$ where $a$
and $\a$ are subgroup and
coset indices and expand the gauge field in components as
\be
A_\pm = A_\pm^a T_a\ ,
\label{e-2-8k}
\ee
where $T^a$ are generators of $H$ appropriately embedded in $G$.
In addition, we introduce the $\dim(H)$ matrix $M(g)$ with elements
\be
M_{ab} = \Tr (t_a g t_b g^{-1}) - \eta_{ab}\ .
\label{e-2-8l}
\ee
Then, the part of the action in \eqn{gwwzw} involving the gauge fields becomes
\be
{k\ov \pi}\int_M \left[\ii (A_-)_a R^a_+ - \ii (A_+)_a L^a_- + A^a_- M_{ab} A^b_+\right]\ .
\ee
Being non-dynamical, the gauge fields $A_\pm$ can be integrated out using their equations of motion,
yielding
\be
A_+^a = - \ii (M^{-1})^{ab} (R_+)_b\ ,\qq A_-^a = \ii (L_-)_b (M^{-1})^{ba} \ .
\label{e-2-2b}
\ee
Substituting back into the action one obtains
\be
 - {k \ov \pi} \int_M  L_{a\mu} (M^{-1})^{ab} R_{b\nu} \partial_- X^\mu \partial_+ X^\nu\ ,
\label{e-2-2c}
\ee
Therefore, the full action assumes the $\s$-model form
\eqn{e-2-9} with \cite{BaSfe1}
\be
G_{\m\n} = G^{(0)}_{\m\n} -  L_{a(\mu} (M^{-1})^{ab} R_{b\nu)}\ ,
\qq B_{\m\n} = B^{(0)}_{\m\n} -  L_{a[\mu} (M^{-1})^{ab} R_{b\nu]}\ ,
\label{e-2-10kl}
\ee
where $G^{(0)}_{\m\n}$ and $B^{(0)}_{\m\n}$ are given by \eqn{e-2-10} and \eqn{e-2-11}.
The process of integrating out the gauge fields induces also a dilaton whose one-loop expression
is given in general by \cite{BaSfe1}
\be
e^{-2\Phi} = \det(M)\ .
\label{diila}
\ee

\no
Of particular interest in this paper are coset models of the type $G^{(1)}_{k_1}\times G^{(2)}_{k_2}/H_{k_1+k_2}$,
with the subgroup $H$ appropriately embedded into the direct product of the groups $G^{(1)}$ and
$G^{(2)}$. The gauged WZW action is
\ba
&& S_{\rm gWZW}(g_1,g_2,A_\pm)  = k_1 I_0(g_1) + k_2 I_0(g_2)
\nonumber\\
&& \phantom{xxxx} + {1\ov \pi} \int_M \Tr \Big[k_1 A_- \del_+ g_1 g_1^{-1} + k_2 A_- \del_+ g_2 g_2^{-1}
 - k_1 A_+ g_1^{-1} \del_- g_1
\label{gwwzw1}\\
&&
\phantom{xxxx} -  k_2 A_+ g_2^{-1} \del_- g_2
+ k_1  A_- g_1 A_+ g_1^{-1} + k_2  A_- g_2 A_+ g_2^{-1}
- (k_1+k_2) A_- A_+\Big]\ ,
\nonumber
\ea
where $g_{1}$ and $g_2$ are elements of the groups $G^{(1)}$ and $G^{(2)}$, respectively.
Then, the previous general formulae are valid provided that one performs the replacements
\ba
&& k L^a_\pm \to k_1 L^{(1)a}_\pm  + k_2 L^{(2)a}_\pm \ ,\qq
k R^a_\pm \to k_1 R^{(1)a}_\pm  + k_2 R^{(2)a}_\pm \ ,
\nonumber\\
&& k M_{ab}(g)\to k_1 M^{(1)}_{ab}(g_1) + k_2 M^{(2)}_{ab}(g_2)\ ,
\ea
where the superscripts $(1)$ and $(2)$ distinguish appropriately the two groups.

\section{Field equations and strategy for solving them}
\setcounter{equation}{0}

The WZW model
has a manifest $G_L\times G_R$ group of isometries  \cite{Witten}.
The gauging and the elimination of the gauge fields leading to a
$\s$-model with background fields \eqn{e-2-10kl} destroys these isometries except those
commuting with the gauge group $H$. This breaking is manifest in explicit constructions of
such backgrounds that can be found in the literature.
For instance, in the simplest such model, one starts with an $SU(2)$ WZW model
and gauges one of the two available abelian symmetries that is either a vector $U(1)_V$ or an
axial $U(1)_A$.
The resulting background \cite{Bardacki, WittenBH}
has a $U(1)$ symmetry corresponding to the $U(1)$ which is not
gauged. This situation is generic even for higher dimensional groups when the gauge group is abelian.
In higher dimensional group manifolds, when a maximal non-abelian subgroup is gauged, there is no isometry left in the resulting background fields. This has been explicitly shown for the
three- and four-dimensional cosets $SO(4)/SO(3)$ and $SO(5)/SO(4)$ and their
non-compact versions \cite{BaSfe1,BaSfe2}. If, instead, there is a
subgroup that commutes with the gauge group, then the corresponding symmetry becomes manifest in the
background geometry. This is the case for the three-dimensional black string geometry based on the
coset $SL(2,\mathbb{R}) \times U(1)/U(1)$ \cite{HorHor} as well as for
the five-dimensional background corresponding to the coset $SU(2,1)/SU(2)$ \cite{Lugo},
which have two commuting isometries.

\no
In physical applications based on the background geometry it is often the case that one deals with
field equations, for instance, for scalar, vector and spinor fields.
Since the background has generically very few, if any, isometries,
solving these equations and extracting physical information is problematic with standard methods, such as separation of variables.
In this section we present a robust method that overcomes this problem, based on the group theoretic structure that underlies the entire construction.

\no
To be concrete, we focus on the scalar field equation, which in a background with metric $G_{\m\n}$ and
dilaton $\Phi$ is of the form
\be
-{1\ov e^{-2\Phi} \sqrt{G}}\ \del_\m e^{-2\Phi} \sqrt{G} G^{\m\n} \del_\n \Psi = E \Psi\ .
\label{lappp}
\ee
Note that the antisymmetric tensor does not enter explicitly, but only implicitly,
being necessary for the consistency of the background.
The presence of the dilaton factor in the measure can be understood in two complementary ways. Firstly,
in the effective actions of low energy string theories the right measure in the string frame is the combination
$e^{-2\Phi} \sqrt{G}$ \cite{callan}. Secondly, we recall that in the Hamiltonian approach \cite{BaSfe4} to
determine the geometry corresponding
to gauged WZW models the scalar equation is nothing but the action of the zero mode of the energy momenum tensor
of the coset construction $G/H$.
This essentially involves the quadratic Casimir operators for the group and the subgroup, where the various
currents are first order differential operators in terms of the elements parametrizing the group element in $G$.
Specifically,
\be
H = {L_A L_A\ov k+g_G} - {L_a L_a\ov k+g_H}\  ,\qq L_A = L^M_A \del_M\ ,
\label{jkf1}
\ee
where $L^M_A$ are the components of the inverse (left-invariant) Maurer--Cartan matrix defined in \eqn{e-2-7}
and $g_G$, $g_H$ are the dual Coxeter numbers for $G$ and $H$.
A completely equivalent expression is found using the right-invariant Maurer--Cartan matrix.
Then, the quadratic Casimirs act as second-order
differential operators on a reduced space spanned by $H$-invariant combinations of these
group element variables \cite{BaSfe3, BaSfe4}.
This effectively reduces it to a differential operator depending on
$\dim(G/H)$ variables.
Identifying it with \eqn{lappp} is possible if and only if the indicated dilaton factor
is present \cite{BaSfe4}.

\no
Due to the fact that $g_G$ and $g_H$ are different the resulting geometry depends non-trivially
on $k$. Our aim in this paper is to determine the eigenfunctions and eigenvalues $\Psi$ and $E$. It turns
out, as will be explained below, that the eigenfunctions do not depend on $k$, so it is enough to restrict
ourselves to the semiclassical limit $k\gg 1$ in which the backgrounds considerably simplify and in
fact are given by \eqn{e-2-10kl} and \eqn{diila}.

\subsection{The $SU(2)$ group manifold and associated cosets}

We first illustrate the basic idea with elementary examples based on the $SU(2)_k$ WZW model.
We use the following parametrization for a group element $g\in SU(2)$
\be
g = e^{ {\ii \ov 2} (\phi_1-\phi_2) \s_3} e^{ {\ii \ov 2} \th \s_2} e^{ {\ii \ov 2} (\phi_1+\phi_2) \s_3}
=
\pmatrix{\cos{\th\ov 2} e^{i\phi_1} &
\sin{\th\ov 2} e^{-i \phi_2} \cr
-\sin{\th\ov 2} e^{i\phi_2} & \cos{\th\ov 2} e^{-i \phi_1} \cr}\ ,
\label{e-2-1kj}
\ee
which is the fundamental $j=1/2$ representation in terms of  angles $(\theta,\phi,\psi)$ and where we have set
$\phi =\phi_1-\phi_2$ and $\psi = \phi_1+\phi_2$.
Inserting into \eqn{e-2-1} and applying the Polyakov--Wiegmann identity, we obtain a $\s$-model with
the $S^3$-metric
\be
ds^2 =k\left({1 \ov 4} d\th^2 + \cos^2 {\th \ov 2} d\phi_1^2 + \sin^2 {\th \ov 2}  d\phi_2^2\right)\ ,
\label{e-2-1b}
\ee
an antisymmetric tensor (not needed for our purposes) and a constant dilaton. In this coordinate system,
the scalar wave equation \eqn{lappp} is readily solved by separation of variables. Using the ansatz
\be
\Psi(\theta,\phi_1,\phi_2) = {1 \ov 2 \pi} \psi(\th) e^{\ii n \phi_1} e^{\ii m \phi_2}\ ,\qq m,n\in \mathbb{Z}\ ,
\label{e-2-1c}
\ee
one obtains the ordinary differential equation
\be
{1 \ov \sin \th} {d \ov d\th} \left(\sin \th {d \psi \ov d \th} \right) -{1\ov 4}
\left( {m^2 \ov \sin^2 {\th \ov 2}} + {n^2 \ov \cos^2 {\th \ov 2}}
\right) \psi = -{kE\ov 4} \psi\ ,
\label{e-2-1d}
\ee
which can be cast by an appropriate transformation into the standard Jacobi equation.
The complete set of normalizable solutions is given by
\be
\psi_{j,m,n}(\th) = % A_{j,m,n}
\left(\sin{\th\ov 2}\right)^{|m|}
\left(\cos{\th\ov 2}\right)^{|n|}
P^{(|m|,|n|)}_{j-{|m|\ov 2}-{|n|\ov 2}}(\cos\th)\ ,\qq
 j-{|m|\ov 2}-{|n|\ov 2}=0,1,\dots \ ,
\label{doo3}
\ee
where  $P_{n}^{(\a,\b)}(x)$ is a Jacobi polynomial.
% and the normalization constant is
%\be
%A^2_{\ell,m,n} = (2\ell+1) {\G(\ell+\ha |m|+\ha |n|+1)
%\G(\ell-\ha |m|-\ha |n|+1)\ov \G(\ell+\ha |m|-\ha |n|+1)
%\G(\ell-\ha |m|+\ha |n|+1)}\ ,
%\label{noorm3}
%\ee
%so that the state is normalized to one with respect to the measure $\ha d\th \sin\th$.
%\be
%\varphi_{\ell,m,n} (\th) = A_{\ell,m,n} P_{\ell - {|m| \ov 2} - {|n| \ov 2}}^{(|m|,|n|)} (\cos \th)\ .
%\label{e-2-1e}
%\ee
%where $P_{n}^{(\a,\b)}(x)$ is a Jacobi polynomial and $A_{\ell,m,n}$ is the normalization constant.
In addition, the spectrum is quantized with
\be
E_j= {4 j(j+1)\ov k}\ .
\label{e-2-1f}
\ee

\no
We want to use these solutions in order to generate the solutions of the wave equation
for the geometric coset $SU(2)/U(1)$ corresponding to the $S^2$-sphere, as well as for the
conformal $SU(2)_k/U(1)$ coset.

\subsubsection{Geometric coset}

The $\s$-model corresponding to the $S^2$-metric is given by
\be
ds^2_{S^2} = (L_{1 \m} L_{1 \n} + L_{2 \m} L_{2 \n})dX^\m dX^\n =  d\th^2 + \sin^2{\th}\ d\phi^2\ ,
\ee
where $L_{i\m}$, $i=1,2$ are components of the Maurer--Cartan forms corresponding to the Pauli-matrices
$\s^i$ and $X^\m=(\th,\phi)$. This reduction of dimensionality gives rise to the geometric coset $SU(2)/U(1)$.

\no
The corresponding wave equation, after substituting an ansatz of the form
\be
\Psi(\th,\phi)={1\ov \sqrt{2\pi}} e^{in\phi} \psi(\th)\ ,\qq
m\in \mathbb{Z}\ ,
\ee
becomes
\be
{1\ov \sin\th}{d\ov d\th}\left(\sin \th {d\Psi\ov d\th}\right)
 -{n^2\ov \sin^2\th} \Psi=- E \Psi\ .
\label{diej31}
\ee
Although it is quite straightforward to solve it, we would like to obtain its solutions indirectly
from the wave equation for $S^3$ given by \eqn{doo3}.
Since the $S^2$-metric is two-dimensional,
if a solution of \eqn{doo3} is to solve \eqn{diej31} as well it
should depend only on two of the original Euler
angles from \eqn{e-2-1kj}, that is on
$\theta$ and $\phi=\phi_1-\phi_2$. This implies that we should
set $m=-n$. We find that
\be
\psi_{j,n}(\th) =  %B_{j,n}
%\sin^{|n|}{\th\ov 2}\cos^{|n|}{\th\ov 2}\
\sin^{|n|}\th\
P^{(|n|,|n|)}_{j-|n|}(\cos\th)\ ,\qq
 j-|n|=0,1,\dots \ .
\label{doyo3}
\ee
%where the normalization constant is obtain from \eqn{noorm3} and is given by
%\be
%B^2_{\ell,n} = {2\ell+1\ov (\ell!)^2} \G(\ell+|n|+1) \G(\ell- |n|+1)\ ,
%\label{ngorm3}
%\ee
%so that the state is, as before, normalized to one with respect to the measure $\ha d\th \sin\th$.
The spectrum is quantized accordingly, as
\be
E_j = j (j +1)\ ,
\label{e-2-1df}
\ee
which corresponds precisely to \eqn{e-2-1f} (ignoring the factor
$4/k$).
We note here that $j$ is an integer.
One can readily verify that \eqn{doyo3} with \eqn{e-2-1df} solve the differential
equation \eqn{diej31}. This is also consistent with the fact that
the differential equation \eqn{diej31} is obtained by setting $m=-n$ in \eqn{e-2-1d}.

\no
An equivalent method, and perhaps more amenable to generalizations in more complicated coset spaces, is to invoke
the subgroup $U(1)_L\times U(1)_R$ of the original global $SU(2)_L\times SU(2)_R$ symmetry of the $S^3$ group
manifold. In terms of the Euler angles in \eqn{e-2-1kj} this correspond to the global shifts
\be
\phi\to \phi + \e_L \ ,\qq \psi\to \psi + \e_R\ .
\ee
The coset reduction requires a singlet under the $U(1)_R$ symmetry and therefore, as for the metric case, the
eigenfunctions of the Laplacian should be truncated to the singlet subsector.
Setting $n=-m$ does precisely that.

\no
The above motivates our basic assertion that solutions of the wave equation
before any reduction is applied contain enough information
to determine the solutions of the same equation in the reduced
geometric coset space.

\subsubsection{Conformal coset}

We next apply the above ideas to the case of the background
corresponding to the gauged WZW model for the $SU(2)_k/U(1)$ CFT given by \cite{Bardacki,WittenBH}
\ba
&& ds^2 =k\left(
{1\ov 4} d\th^2 + \tan^2{\th\ov 2} \ d\varphi^2\right)\ , \qq 0\leqslant \th<\pi\ ,
\quad 0\leqslant \varphi<2\pi\ ,
 \nonumber\\
&& e^{-2 \Phi} =e^{-2 \Phi_0} \cos^2{\th\ov 2}\ ,
\label{su2cft}
\ea
where $\Phi_0$ is a constant.
This background is obtained by gauging, in the sense described in section 2, the
axial subgroup $U(1)_A$ of the original global $SU(2)_L\times SU(2)_R$ symmetry of the $SU(2)$ WZW.
In terms of the Euler angles in \eqn{e-2-1kj} it correspond to the shifts
\be
\phi\to \phi + \e \ ,\qq \psi\to \psi + \e\ .
\label{ssyv}
\ee
For the wave equation consider an ansatz of the form
\be
\Psi(\th,\varphi)={1\ov \sqrt{2\pi}} e^{im\varphi} \psi(\th)\ ,\qq
m\in \mathbb{Z}\ .
\ee
Then the amplitude obeys
\be
{1\ov \sin\th}{d\ov d\th}\left(\sin \th {d\psi\ov d\th}\right)
-{m^2\ov 4} \cot^2{\th\ov 2}\psi=-{k E\ov 4}\psi\ .
\label{ditf1}
\ee
\no
The coset reduction should produce a singlet under the $U(1)_V$ symmetry.
As before, in order to project the eigenfunctions \eqn{doo3}
to the singlet subsector of $U(1)_V$ we should choose appropriately the eigenvalues.
In this case the appropriate choice is to set
$n=0$.
Then the solution \eqn{e-2-1c} depends only on $\phi_2=\ha(\psi-\phi)$, which, according
to \eqn{ssyv}, is a gauge invariant combination.
The solution is given by
\be
\psi_{j,m}(\th) =  %\sqrt{2j+1}
\left(\sin{\th\ov 2}\right)^{|m|}\
P^{(|m|,0)}_{j-{|m|\ov 2}}(\cos\th)\ ,\qq
 j-{|m|\ov 2}=0,1,\dots \ .
\label{djkoo3}
\ee
%so that the state, as always, is normalized to one with respect to the measure $\ha d\th \sin\th$.
Note that now $j$ is a half integer.
Substituting the solution into \eqn{ditf1}
we find that the spectrum is quantized accordingly as
\be
E_{j,m}= {4 j(j+1)\ov k}-{m^2\ov k}  \ .
\label{hlspq}
\ee
This is consistent with the fact that \eqn{ditf1} is obtained from \eqn{e-2-1d} by setting $n=0$ only after we shift the energy as above. In addition, it is consistent with the CFT result
in the large $k$-limit, as can be seen from \eqn{jkf1} above.

\subsection{The general algorithm}

The construction of eigenstates for a general conformal coset can be
done using the group structure of the model. In spite of the absence of
any generic isometries, states can still be found explicitly.

\no
The starting point is the set of eigenstates of the Laplacian on the
original group manifold $G$, which need not be simple.
If $R$ is an irreducible representation (irrep) of $G$,
then the set of all matrix elements of $g \in G$ in all irreps
$R$ constitute a complete set of eigenstates for the Laplacian on the
group manifold.

\no
To review the above fact, denote by $R_{\alpha \beta}(g)$
the matrix elements of $g$ in the irrep $R$. They obey the
group property
\be
R_{\alpha \beta} (g_1 g_2 ) = \sum_\gamma
R_{\alpha \gamma} (g_1 ) R_{\gamma \beta} (g_2 )\ ,
\ee
where we assume that we work in a basis in which the invariant Killing metric is the identity.
The generators of left transformations on the group manifold $L^A$
act on $R(g)$ as
%\begin{eqnarray}
%(1+i\varepsilon ) L^A R_{\alpha \beta} (g) &=& R_{\alpha \beta}
%((1+i\varepsilon T^A) g) = \sum_\gamma R_{\alpha \gamma}
%(1+i\varepsilon T^A)R_{\gamma \beta} (g) \nonumber \\
%&=& R_{\alpha \beta} (g) + i\varepsilon
%R_{\alpha \gamma}^A R_{\gamma \beta} (g)\ ,
%\end{eqnarray}
\be
L^A R_{\alpha \beta} (g) = \sum_\gamma
R_{\alpha \gamma}^A R_{\gamma \beta} (g)\ ,
\ee
where $R^A = R(T^A )$ is the $A$-th generator of $G$ in the
$R$ irrep.
The scalar field equation on $G$ is $H \psi = E \psi$,
where $H$ is essentially the (negative) Laplacian expressed as the quadratic sum
of $L^A$ and acting on $R(g)$ as
\be
HR_{\alpha \beta} (g)= {\sum_A (L^A )^2\ov k+g_G} R_{\alpha \beta} (g)
%\sum_{A;\gamma, \delta} R_{\alpha \gamma}^A R_{\gamma \delta}^A R_{\delta \beta} (g)
= {1\ov k+ g_{G}} \sum_{A;\gamma} (R^A )_{\alpha \gamma}^2 R_{\gamma \beta} (g)\ .
\ee
The sum $\sum_A (R^A )^2$ is the quadratic Casimir of $G$ and,
in the irrep $R$, it is proportional to the identity matrix
$C_2 (R) \delta_{\alpha \gamma}$, giving
\be
H R_{\alpha \beta} (g) ={ C_2 (R)\ov k + g_G} R_{\alpha \beta} (g)\ ,
\ee
%where $g_G$ is the dual Coxeter number for $G$.
So we obtain $\dim(R)$ degenerate eigenstates corresponding to the
eigenvalue $E(R) = C_2 (R)/(k + g_G )$. A similar arguments works in terms
of the generators of right transformations $R^A$ since the corresponding eigenvalues of the quadratic
Casimir are the same on the above states.

\no
Under a global left- and right-rotation $g \to g_{_L} g g_{_R}^{-1}$ the above
degenerate eigenstates transform as
\be
R_{\alpha \beta} (g_{_L} g g_{_R}^{-1} ) = \sum_{\gamma, \delta}
R_{\alpha \gamma} (g_{_L} )
R_{\gamma \delta} (g) R_{\delta \beta}^{-1} (g_{_R} )\ .
\ee
So the left index in $R_{\alpha \beta} (g)$ transforms in the irrep
$R$ under left rotations and the right index transforms in the
conjugate representation $\bar R$ under right rotations of $g$.

\no
To identify the eigenstates of the Laplacian on the conformal
coset manifold $G_k/H_k$, with $H$ a subgroup of $G$, we need to find
linear combinations of the states $R_{\alpha \beta} (g)$ that
are singlets under the transformation $g \to h g h^{-1}$, $h \in H$.
The irrep $R$ of $G$ is generically reducible under $H$ and
decomposes into a direct sum of irreps $r_i$ of $H$.
So under $g \to h g h^{-1}$ the states
$R_{\alpha \beta} (g)$ transform in the representation
\be
( r_1 \oplus r_2 \oplus \cdots ) \otimes ( {\bar r}_1 \oplus {\bar r}_2
\oplus \cdots )\ ,
\ee
with each index in $R_{\alpha \beta} (g)$ transforming in the
corresponding term of the above direct product.

\no
We form states
invariant under the coset transformation by identifying the
singlets in the above direct product. Contracting each pair
$(r_i , {\bar r}_i)$
we can form one singlet. So we obtain one eigenstate of the Laplacian
on the coset manifold for each distinct irrep $r_i$ of $H$ included
in the irrep $R$ of $G$. (If an irrep $r_i$ is included $n$ times
in $R$ then we will get $n^2$ degenerate eigenstates.)
Denoting by $C_\alpha^a (R,r_i )$ the
decomposition coefficients projecting the state $\alpha$ of $R$ into the state
$a$ of $r_i$, the corresponding $G_k/H_k$ eigenstate will be
\be
\psi_{R,r_i} (g) = \sum_{a; \alpha, \beta} C_{\alpha}^{a} (R,r_i )
C_{\beta}^{a} (R, r_i ) R_{\alpha \beta} (g)\ ,
\ee
with eigenvalue, according to \eqn{jkf1}, given by the coset Sugawara expression
\be
E(R,r_i) = {C_2 (R)\ov k+g_G} -{ C_2 (r_i )\ov k+g_H}\ .
\ee

\no
The above analysis is completely general. In cases of direct product groups of the type
$G^{(1)}_{k_1} \times G^{(2)}_{k_2}/H_{k_1+k_2}$,
where $H$ is a diagonal subgroup appropriately embedded in the direct product,
the configuration space is parametrized
by the two group elements ($g_1 ,g_2)$ modulo the identification
\be
( g_1 , g_2 ) \sim ( h g_1 h^{-1} , hg_2 h^{-1} )
\ ,\qq  g_i\in G^{(i)}\ ,\quad i =1,2\ ,\quad h \in H\ .
\ee
The irreps of $G^{(1)} \times G^{(2)}$ are direct products of two irreps of the $G^{(i)}$s,
$R^{(1)} \times R^{(2)}$. Therefore, the eigenstates of the Laplacian on the full group
manifold are
\be
R^{(1)}_{\alpha \beta} (g_1) R^{(2)}_{\m\n} (g_2)\ ,
\ee
with eigenvalues
\be
E(R) = \frac{C^{(1)}_2 (R)}{k_1 + g_{G^{(1)}}} + \frac{C^{(2)}_2 (R)}{k_2+ g_{G^{(2)}}}\ .
\ee
Under a vector $H$-transformation the above states transform in
the representation
\be
( R^{(1)} \times R^{(2)} ) \times ({\bar R}^{(1)} \times {\bar R}^{(2)} )\ .
\ee
Decomposing $R^{(1)} \times R^{(2)} $ into irreps $r_i$ of $H$ and
denoting by $C_{\alpha \m}^{a} (R^{(1)},R^{(2)};r_i )$ the
Clebsch--Gordan coefficient
projecting the state $\alpha$ of $R^{(1)}$ and the state $\m$ of $R^{(2)}$ into the state
$a$ of $r_i$, we construct coset eigenstates as
\be
\psi_{R^{(1)},R^{(2)};r_i} (g_1,g_2)  = \sum_{a;\alpha,\beta,\mu,\nu}
C_{\alpha \mu}^{a} (R^{(1)},R^{(2)};r_i)
C_{\beta \nu}^{a} (R^{(1)},R^{(2)};r_i )
R^{(1)}_{\alpha \beta} (g_1) R^{(2)}_{\mu \nu} (g_2)\ ,
\label{fhhqo}
\ee
with eigenvalues
\be
E(R^{(1)},R^{(2)};r_i) = \frac{C_2 (R^{(1)})}{k_1+ g_{G^{(1)}}} + \frac{C_2 (R^{(2)} )}{k_2+ g_{G^{(2)}}} -
\frac{C_2 (r_i )}{k_1 + k_2+ g_H}\ .
\ee
This construction can be carried out explicitly whenever the
expressions for the representations $R(g)$ and the Clebsch--Gordan
coefficients $C_{\alpha \mu}^{a} (R^{(1)},R^{(2)} ;r_i )$ of $r_i \in R^{(1)} \times R^{(2)}$
are known.

\no
The above eigenstates should constitute a complete orthogonal set of solutions of the wave equation
for the background of the corresponding  coset CFT. For our example this will be verified below.

\no
In addition, it is obvious from the above construction
that the states of the theory do not depend on the level
$k$ (or $k_1$ and $k_2$ in the case of direct product).
The levels appear non-trivially only in the eigenvalues.
Consequently, to simplify the upcoming discussion,
we will only keep the semiclassical
expressions for the background fields in which the Coxeter numbers are ignored. The full dependence of the eigenvalues on the renormalized
levels can easily be restored.

\section{A non-trivial example: $SU(2)_{k_1} \times SU(2)_{k_2} / SU(2)_{k_1+k_2}$}

\setcounter{equation}{0}

As a non-trivial test of our general idea, consider the coset
$SU(2)_{k_1} \times SU(2)_{k_2} / SU(2)_{k_1+k_2}$, for which the general gauged WZW action
\eqn{gwwzw1} for direct product groups can be used.

\subsection{The background geometry}

For our purposes, it will be most convenient to adopt the parametrization for the associated group elements
of the fundamental representation as
\be
g_1 =
\left(
  \begin{array}{cc}
    \a_0 + i \a_3 & \a_2 + i \a_1 \\
    -\a_2+i \a_1 & \a_0 - i \a_3 \\
  \end{array}
\right)\ ,\qq g_2 =
\left(
  \begin{array}{cc}
    \b_0 + i \b_3 & \b_2 + i \b_1 \\
    -\b_2+i \b_1 & \b_0 - i \b_3 \\
  \end{array}
\right)\ ,
\label{e-2-8a}
\ee
where from unitarity
\be
\a_0^2 + \vec{\a}^2 = 1\ , \qq \b_0^2 + \vec{\b}^2 = 1\ .
\label{e-2-8b}
\ee

\no
Consider first the corresponding WZW action.
Inserting the above parametrization into the WZW part of the action and using the constraints \eqn{e-2-8b},
it is easily seen that it leads to a $\s$-model with metric given by $ds^2_{(1)} + ds^2_{(2)}$
where
\be
ds^2_{(1)} = %{4 k_1 \ov 1- {\vec \a}^2}
%\left[(1-\a_2^2-\a_3^2) d\a_1^2 + 2 \a_1 a_2 d\a_1 d\a_2 + {\rm cyclic\ in\ 1,2,3}\right] \ ,
4 k_1 \left(\d_{ij} + {\a_i\a_j\ov \a_0^2}\right) d\a_i d\a_j \ ,
\label{e-2-8c}
\ee
whereas the antisymmetric tensor has field strength
\be
H^{(1)} = {8 k_1\ov \a_0} d\a_1 \wedge d\a_2 \wedge  d\a_3\ .
\label{ttros}
\ee
Similar expressions hold for $ds^2_{(2)}$ and $H^{(2)}$.

\no
Next we gauge the diagonal $SU(2)$ subgroup of the full $SU(2) \times SU(2)$ group.
It turns out that, in their infinitesimal form, the left group of transformations act as
\be
\d_L \a_0 = -\ha \e_L^{(i)} \a_i \ ,\qq \d_L \a_i = \ha \a_0 \e_L^{(i)} +\ha  \e_{ijk} \a_j \e_L^{(k)}\ .
\ee
For the right group of transformations we have instead
\be
\d_R \a_0 = -\ha \e_R^{(i)} \a_i \ ,\qq \d_R \a_i = \ha \a_0 \e_R^{(i)} -\ha  \e_{ijk} \a_j \e_R^{(k)}\ .
\ee
Notice that the combined transformation $\d_L + \d_R\equiv \d $
does not close into a group unless $\e_L^{(i)} = -\e_R^{(i)} = \e^{(i)}$, that is we should consider the vector
gauging, a situation that is generic when the gauge group is non-abelian (for possible generalizations
in asymmetric cosets see \cite{BaSfe1,obers,quella}).
Then we have that
\be
\d \a_i =  \e_{ijk} \a_j \e_k\ ,\qq \d \b_i =  \e_{ijk} \b_j \e_k\ ,
\ee
implying that $\vec \a$ and $\vec \b$ indeed transform as vectors.
On general grounds,
the background is expected to depend only on invariants of these
three-vectors. They can be chosen to be the three combinations
\ba
&& \a = |\vec{\a}|\ ,\qq
\b = |\vec{\b}|\ ,\qq \g = \vec \a\cdot \vec \b ,
\nonumber\\
&& 0\leqslant \a,\b\leqslant 1\ ,\qq |\g|\leqslant \a\b\ .
\label{abfg}
\ea
Alternatively, we may fix the gauge as
\be
a_2 = \a_3 = \b_3 = 0 \
\label{gff0}
\ee
and then perform the computation, which is greatly simplified.
For instance, when the gauge is fixed the torsion in \eqn{ttros}
vanishes from the very beginning.
At the end we may pass to the gauge invariant combinations
by setting
\be
\a_1= \a\ ,\qq \b_1={\g \ov \a}\ ,\qq \b_2 = \sqrt{\b^2 - {\g^2/\a^2}}\ .
\label{gff1}
\ee
which follow by inserting the gauge-fixed expressions into \eqn{abfg}.
It is convenient to
introduce the ratio
\be
r = {k_2 \ov k_1} \ .
\label{e-2-9q}
\ee
Then, by following the general procedure and after a tedious computation, we obtain a
$\s$-model with metric\footnote{Since $|\g| \leqslant \a\b$,
a more appropriate global parametrization would be $\g =
\a \b \cos \varphi$, with
$0\leqslant \varphi \leqslant \pi$. This global parametrization of $\g$ will be useful in checking the
orthogonality properties of the eigenstates of the scalar equation as we shall see at the
end of subsection 4.2.}
\ba
ds^2 &=& {k_1 + k_2 \ov (1-\a_0^2 ) (1-\b_0^2 ) - \g^2}
\bigl( \D_{\a\a} d\a_0^2 + \D_{\b\b} d\b_0^2 + \D_{\g\g} d\g^2 \nonumber\\
&& + 2 \D_{\a\b} d\a_0 d\b_0 + 2 \D_{\a\g} d\a_0 d\g + 2 \D_{\b\g} d\b_0 d\g \bigr) \
\label{e-2-10q}
\ea
with
\ba
&&\D_{\a\a} = {(1+r)^2 -r(2+r) \b_0^2 \ov r(1+r)^2}\ ,
\qq \D_{\b\b} = {(1+r^{-1})^2 -r^{-1}(2+r^{-1}) \a_0^2\ov r^{-1} (1+r^{-1})^2}\ ,
\nonumber\\
&&\D_{\g\g} = {1 \ov 2 + r +r^{-1}}\ ,
\qq \phantom{xx}
\D_{\a\b} = \g + {\a_0 \b_0 \ov 2+r+r^{-1}} \ ,
\label{e-2-11q}\\
&&\D_{\a\g} = -{ \b_0 \ov (1+r)^2}\ ,
\qq \phantom{xx} \D_{\b\g} = -{ \a_0 \ov (1+r^{-1})^2} \ .
\nonumber
\ea
The antisymmetric tensor is zero and the dilaton reads (up to a constant factor)
\be
e^{-2 \Phi} =   (1-\a_0^2 ) (1-\b_0^2 ) - \g^2 \ .
\label{e-2-12}
\ee
The background is manifestly invariant under the interchange of $\a_0$ and $\b_0$ and a simultaneous inversion
of the parameter $r$. This symmetry simply interchanges the two $SU(2)$s that we gauge.

\no
We have checked that the above background indeed solves the one-loop $\b$-function
for conformal invariance, as it should.
Since the subgroup of $SU(2)\times SU(2)$ that has been gauged is the maximal one, namely
the diagonal $SU(2)$ subgroup, the above metric is not expected to have any isometries and in fact it
does not appear to have any (although we have not checked that explicitly). We note that, even if there is an
accidental isometry, it is useful only if it can be made manifest.
That would necessarily involve a change of variables
after which the global properties of the new coordinates would be less transparent.

\no
We also note that the background fields for this coset
have also been worked out in \cite{Cre} (and implicitly in \cite{BaSfe1, BaSfe3} which dealt with the coset
$SO(4)/SO(3)$). Due to the different parametrization used
in these references we found it more efficient to proceed with an independent derivation.

\subsection{Solving the wave equation}

Clearly, the complexity of the metric \eqn{e-2-11q}, and in particular the apparent lack of isometries,
makes the scalar equation \eqn{lappp} very difficult to solve. Any attempt to integrate it using
conventional methods, such as separation of variables, would be hopeless. We will, however, present the full solution according to our previous general discussion, by first introducing some notation.

\no
Consider the fundamental representation of $GL(2,\mathbb{R})$
\be
R^{1/2}= \left(
           \begin{array}{cc}
             a & b \\
            c & d \\
           \end{array}
         \right) \ .
\ee
A general representation $R^j$ has matrix elements \cite{Vilenkin}
\ba
R^j_{m_1,m_2} (a,b,c,d) =
\sum_k A^{j}_{m_1,m_2,k}\
  a^{j-m_1-k} d^{j+m_2-k} b^k c^{k+m_1-m_2}\ ,
\ea
where
\be
A^{j}_{m_1,m_2,k}
 = {\sqrt{(j+m_1)!(j-m_1)! (j+m_2)!(j-m_2)!}\ov k! (j-m_1-k)! (j+m_2-k)! (k+m_1-m_2)!}\ .
\ee
The summation over $k$ extends to all values for which the factorials have non-negative arguments.

\no
For the group $SU(2)$ that we are interested in particular, the fundamental representation
$R^{1/2}$ is identified with
\eqn{e-2-1kj}. In addition, $j$ is a half-integer and $m_1,m_2 = -j,-j+1,\dots, j$. Then,
the above sum is finite, ranging
between the extreme values ${\rm max}(0,m_2-m_1)$ and ${\rm min}(j+m_2,j-m_1)$.
For the $SU(2)$ case it is customary to use the notation $D^j_{m_1,m_2}(\phi,\th,\psi)$,
which are the so-call $D$-functions. Using the parametrization \eqn{e-2-1kj} these can be expressed as
\be
D^j_{m_1,m_2}(\phi,\th,\psi) = e^{-i(m_1 \phi + m_2 \psi)} d^j_{m_1,m_2}(\th)\ ,
\ee
where the Wigner's $d$-matrix can be written in terms of Jacobi polynomials.
Introducing for notational convenience the integers
\be
m= m_1-m_2\ ,\qq n= m_1+m_2\ ,
\ee
one proves that
\ba
&& d^j_{m_1,m_2}(\th) =   (-1)^{\ha (|m| + m)}\
 \sqrt{ (j+|m|/2+|n|/2)!\
(j-|m|/2-|n|/2)!\ov (j-|m|/2+|n|/2)! \ (j+|m|/2-|n|/2)!}
\nonumber\\
&&\phantom{xxxxxxxxx}
\times \ \left(\sin{\th\ov 2}\right)^{|m|}
\left(\cos{\th\ov 2}\right)^{|n|}
 P_{j-|m|/2-|n|/2}^{|m|,|n|}(\cos\th)\ ,
\label{wigne}
\ea
where the first factor introduces a minus sign whenever $m$ is an odd positive integer.
We note that in this way we recover \eqn{doo3} found previously by solving directly the
wave equation.

\no
Since we have two $SU(2)$ factors we label the corresponding representations $R^{j_1} (g_1)$ and $R^{j_2} (g_2)$,
where $g_1$ and $g_2$ are the fundamental representations parametrized as in \eqn{e-2-8a}.
Then the general state is, according to \eqn{fhhqo}
\be
\Psi^j_{j_1,j_2} = \sum_{m=-j}^j\  \sum_{m_1,n_1=-j_1}^{j_1} \ \sum_{m_2,n_2=-j_2}^{j_2}
C^{j,m}_{j_1,m_1,j_2,m_2}\ C^{j,m}_{j_1,n_1,j_2,n_2}\  R^{j_1}_{m_1,n_1}(g_1)\
R^{j_2}_{m_2,n_2}(g_2)\ ,
\label{pjhr}
\ee
where %$C^{j_1,j_2,j}_{m_1,m_2,m}$'s
$C^{j,m}_{j_1,m_1,j_2,m_2}$ are the Clebsch--Gordan coefficients for a state $|j,m\rangle $
in the diagonal $SU(2)_L$ composed from states $|j_1,m_1\rangle |j_2,m_2\rangle$ in $SU(2)_L\times SU(2)_L$.
Similarly, $C^{j,m}_{j_1,n_1,j_2,n_2}$ are the Clebsch--Gordan coefficients for a state $|j,m\rangle $
in the diagonal $SU(2)_R$ composed from states $|j_1,n_1\rangle |j_2,n_2\rangle$ in $SU(2)_R\times SU(2)_R$.
The sum ensures that a singlet of the diagonal $SU(2)_L\times SU(2)_R$ is formed.
Of course in the above summations we should also ensure that
\be
m=m_1+m_2=n_1+n_2\ ,\qq |j_1-j_2|\leqslant j \leqslant j_1+j_2\ .
\ee
%\be
%\Psi_{j_1,j_2,j} =\sum_{m_1,n_=-j_1}^{j_1} \sum_{m_2,n_2=-j_2}^{j_2} \d_{m_1+m_2,n_1+n_2}
%C^{j_1,j_2,j}_{m_1,m_2,m_1+m_2}\ C^{j_1,j_2,j}_{n_1,n_2,n_1+n_2}\  R^{j_1}_{m_1,n_1}\
%\tilde R^{j_2}_{m_2,n_2}\ , \quad |j_1-j_2|\leqslant j \leqslant j_1+j_2\ .
%\ee
Taking these into account we can reduce the number of summations in \eqn{pjhr} and
write
%\ba
%&& \Psi^j_{j_1,j_2} = \sum_{m=-j}^j\ \sum_{m_2,n_2}
%C^{j,m}_{j_1,m-m_2,j_2,m_2}\ C^{j,m}_{j_1,m-n_2,j_2,n_2}\  R^{j_1}_{m-m_2,m-n_2}(g_1)\
%R^{j_2}_{m_2,n_2}(g_2)\ ,
%\nonumber\\
%&&  \phantom{xxxxxxxxxxxx} {\rm where}\
%-{\rm min}(j_1-m,j_2)\leqslant m_2,n_2 \leqslant {\rm min}(j_2,m+j_1)\ .
%\label{pjhr1}
%\ea
\ba
&& \Psi^j_{j_1,j_2} = \sum_{m} \sum_{m_2,n_2=-j_2}^{j_2}
C^{j,m}_{j_1,m-m_2,j_2,m_2}\ C^{j,m}_{j_1,m-n_2,j_2,n_2}\  R^{j_1}_{m-m_2,m-n_2}(g_1)\
R^{j_2}_{m_2,n_2}(g_2)\ ,
\nonumber\\
&&  \phantom{xxxxx} {\rm where}\
-{\rm min}(j_1-m_2,j_1-n_2,j)\leqslant m \leqslant {\rm min}(j_1+m_2,j_1+n_2,j)\ .
\label{pjhr1}
\ea
In explicitly evaluating this we will use the gauge fixing \eqn{gff0}
followed by \eqn{gff1}.
In addition, we will use the explicit formulae for the Clebsch--Gordan
coefficients (see, e.g., page 3 of \cite{RacahII})
\ba
&& C^{j,m}_{j_1,m_1,j-n_1,m-m_1} = \sum_k (-1)^k
\left( 2 j +1\ov 2 j+1 + j_1 -n_1 \right)^{1/2}
\nonumber\\
&& \times\ {[(j_1-n_1)!(j_1+n_1)!(j_1+m_1)!(j_1-m_1)!]^{1/2}\ov k!
(j_1-m_1-k)! (n_1+m_1+k)!(j_1-n_1-k)!}
\\
&& \times\ \left((2 j -j_1-n_1)!(j+m-n_1-m_1)! (j-m-n_1+m_1)!
(j+m)!(j-m)! \ov (2 j + j_1-n_1)! [(j+m-n_1-m_1-k)!
(j-m-j_1+m_1+k)!]^2\right)^{1/2}\ ,
\nonumber
\ea
where, as before, the summation extends
to all values for which the arguments of the factorials are
non-negative. Note that we have introduced the half integer $n_1$ to
parametrize the deviation of the spin $j_2$ from $j$, which one
easily sees that it obeys $|n_1|\leqslant j_1$.

\no
The state \eqn{pjhr1} should have an eigenvalue equal to
\be
E^j_{j_1,j_2} = {j_1(j_1+1)\ov k_1} + {j_2(j_2+1)\ov k_2} - {j(j+1)\ov k_1+k_2}\ .
\label{enrjjj}
\ee
Given a pair of values for ($j_1,j_2$) there are $2 j_{\rm min}+1$ values for $j$,
where $j_{\rm min}$ is the minimum
of the $j_i$'s. The eigenvalues corresponding to these are in general non-degenerate,
in resonance with the absense of isometries of the background.

\no
A particularly interesting special case is the one in which one of the spins is zero.
Choosing, for instance, $j_2=0$ and thus $j_1=j$, we have that
\be
\Psi^j_{j,0} = \sum_{m=-j}^j R^j_{m,m}(g_1)\ .
\ee
This is nothing but the character of the representation, that is
\be
\Psi^j_{j,0} = {\sin (j+\ha)\th \ov \sin \th/2}\ ,\qq  \cos{\th \ov 2} = \a_0
\label{psioo}
\ee
where in relating the angle $\th$ and $\a_0$ we used the group parametrization in terms of Euler angles
\eqn{e-2-1kj} together with \eqn{gff0} and \eqn{gff1}. We can also express
\eqn{psioo} directly in terms of $\a_0$ as
\ba
\Psi^j_{j,0} & = & \sum_{m=0}^{[j]} (-1)^m %\left(\begin{array}{c}
                                         %2 j+1 \\
                                         %2 m+1 \end{array}\right)
{(2j+1)!\ov (2 m+1)!\ (2j-2 m)!}\
\a_0^{2(j-m)} (1-\a_0^2)^{m}
\nonumber\\
& = & 2^{2 j} \a_0^{2 j} - 2^{2j-2} (2 j-1) \a_0^{2 j-2} + \cdots = U_{2j}(\a_0)\ ,
\label{exppl}
\ea
where in the last step we used the standard notation $U_{2j}(\a_0)$
for the Chebyshev polynomials of the
2nd kind.

\no
Similarly, for the state with $j_1=0$ and $j_2=j$  we have that
\be
\Psi^j_{0,j} = \sum_{m=-j}^j R^j_{m,m}(g_2) = U_{2j}(\b_0)\ .
\label{expp2}
\ee

\no
We present below some further explicit examples:

\no
For $(j_1,j_2)=(1/2,0)$:
\be
\Psi^{1/2}_{1/2,0} = 2 \a_0 \ .
\ee
For $(j_1,j_2)=(1/2,1/2)$:
\ba
&& \Psi^{0}_{1/2,1/2} =  \a_0 \b_0 + \g \ ,
\nonumber\\
&&\Psi^{1}_{1/2,1/2} =  3\a_0 \b_0 - \g \ .
\ea
For $(j_1,j_2)=(1,0)$:
\be
\Psi^{1}_{1,0} = 4 \a_0^2 -1 \ .
\ee
For $(j_1,j_2)=(1,1/2)$:
\ba
&&\Psi^{1/2}_{1,1/2} = {2\ov 3} \left[(4 \a_0^2-1) \b_0 + 4 \a_0 \g \right]\ ,
\nonumber\\
&&\Psi^{3/2}_{1,1/2} = {4\ov 3} \left[(4 \a_0^2-1) \b_0 - 2 \a_0 \g \right]\ .
\ea
For $(j_1,j_2)=(1,1)$:
\ba
&&\Psi^{0}_{1,1} = {1\ov 3} \left[4( \a_0 \b_0 +  \g)^2-1 \right]\ ,
\nonumber\\
&&\Psi^{1}_{1,1} =
6 \a_0^2 \b_0^2 + 4 \a_0 \b_0 \g - 2 \g^2 -2 (\a_0^2+\b_0^2)+1\ ,
\\
&&\Psi^{2}_{1,1} = {1\ov 3} \left[26 \a_0^2 \b_0^2 -20 \a_0 \b_0\g
+ 2 \g^2 -6 (\a_0^2+\b_0^2)+1\right]\ .
\nonumber
\ea
For $(j_1,j_2)=(3/2,0)$:
\be
\Psi^{3/2}_{3/2,0} = 4\a_0( 2\a_0^2 -1)\ .
\ee
For $(j_1,j_2)=(3/2,1/2)$:
\ba
&&\Psi^{1}_{3/2,1/2} = 3 \a_0 \b_0 (2 \a_0^2-1) + (6\a_0^2-1) \g\ ,
\nonumber\\
&&\Psi^{2}_{3/2,3/2} = 5 \a_0 \b_0 (2 \a_0^2-1) - (6\a_0^2-1) \g\ .
\ea
For $(j_1,j_2)=(3/2,1)$:
\ba
&&\Psi^{1/2}_{3/2,1} = {2\ov 3} \left[6 \a_0^3 \b_0^2 - \a_0(2 \b_0^2+1) +2 \b_0(6 \a_0^2-1)\g
+ 6 \a_0 \g^2\right]\ ,
\nonumber\\
&&\Psi^{3/2}_{3/2,1} =
{4\ov 15} \left[6 \a_0^3 (8 \b_0^2-3) - \a_0 (28 \b_0^2-13) + 4 \b_0(6 \a_0^2-1)\g
-24 \a_0 \g^2\right]\ ,
\\
&&\Psi^{5/2}_{3/2,1} =
{2\ov 5} \left[\a_0^3 (38 \b_0^2-8) - 3  \a_0 (6 \b_0^2-1) - 6 \b_0(6 \a_0^2-1)\g
+6  \a_0 \g^2\right]\ .
\nonumber
\ea
Obviously, after a while the expressions become quite complicated. In all cases we have verified that
the indicated functions indeed solve the wave equation with the appropriate eigenvalues given by \eqn{enrjjj}.
All solutions of the form $\Psi^j_{j,0}$ agree with the general result \eqn{exppl}.
As stressed before, the levels $k_1$ and $k_2$ do not appear in the eigenfunctions, not even through
their ratio $r$. They only appear in the associated eigenvalues.

\no
We have checked that the above eigenstates are indeed orthogonal for different values of the
triad $(j_1,j_2,j)$ with respect to the measure
\be
e^{-2\Phi} \sqrt{G}\ d\a_0 \wedge d\b_0 \wedge d\g\ \sim\ \sqrt{(1-\a_0^2 )
(1-\b_0^2 )} \cos\varphi\ d\a_0
\wedge d\b_0 \wedge d\varphi\ ,
\ee
where we have used the global parametrization for $\g$ as described in footnote 1 above.

\subsection{High spin limit and the corresponding effective geometry}

Of particular interest is the large spin behaviour. We assume that one of the spins
becomes large, whereas the other one is kept finite. For instance,
consider
\be
j_1\gg 1\ , \quad  j_2 = {\rm finite} \quad \Longrightarrow \quad j\gg 1\ ,
\label{larspin2}
\ee
In this limit, the eigenvalues \eqn{enrjjj} become infinite unless the level $k_1$ becomes
large as well, but in a way proportional to $j$. Specifically, let
\be
j_1= j-n \ ,\qq k_1 = {k_2\ov \d}\\ j ,
\label{lkjl}
\ee
where $n$ is a half-integer and $\d$ a positive real parameter.
Then
\be E_{j_2,n,\d}= \lim_{j\to
\infty} E^j_{j_1,j_2} = {j_2(j_2+1)\ov k_2} +  {\d - 2 n\ov k_2}\ \d\ .
\label{eiginfi}
\ee
Taking the level $k_1\to \infty$ has implications for the geometry that can support the
corresponding states with infinite spin. It is straightforward to show that in order for the background
to have a good limiting behaviour one should focus on part of the manifold. Consider
focusing near $\a_0=0$.
Accordingly, we define a new variable and take the limit
\be
\a_0 = r \l \ ,\qq r\to 0\ ,
\ee
where $\l$ is the new uncompactified coordinate to be used instead of $\a_0$.
In this limit the product $j \a_0$ remains finite.
We obtain for the metric the expression
\be
ds^2 =   {k_2\ov 1-\b_0^2 -\g^2}\Big(d\l^2 + d\b_0^2 + d\g^2
+ 2 \g \ d\l d\b_0 - 2 \b_0\ d\l d\g \Big)\ ,
\label{kli1}
\ee
whereas the dilaton becomes
\be
e^{-2\Phi} = 1 - \b_0^2-\g^2\ .
\label{kli2}
\ee
This background can be written in a simpler form by performing the
transformation
\be
\g = \sin\th \cos(\varphi + \l) \ ,\qq \b_0 = \sin\th \sin(\varphi + \l)\ ,
\label{jdperi}
\ee
yielding the result
\be
ds^2 =k_2( d\l^2 + d\th^2 + \tan^2\th\ d\varphi^2)\ ,\qq e^{-2\Phi} = \cos^2\th\ .
\label{hgeq}
\ee
This is locally the background for the exact CFT $SU(2)_{k_2}/U(1) \times \mathbb{R}$, whose
non-trivial first factor has been used before in \eqn{su2cft}.

\no
Consider the states in \eqn{psioo} for which $j_2=0$ and therefore $j_1=j$.
Then we have that (up to a constant)
\be
\Psi = \lim_{j\to \infty} \Psi^j_{j,0} = \sin{2 \d \l}\ .
\label{hfho}
\ee
This indeed solves the scalar field equation \eqn{lappp} with eigenvalue $\d^2/k_2$
in accordance with $E_{0,0,\d}$ in \eqn{eiginfi}.
Also note the states $\Psi^j_{0,j}$, given by \eqn{expp2},
which are insensitive to the high spin limit that we consider in which only $j_1$ becomes large.
Nevertheless they should solve the scalar wave equation
with eigenvalue $j(j+1)/k_2$. We have checked that this is indeed the case, in agreement also with
$E_{j,0,0}$ in \eqn{eiginfi}.

\no
The spectrum corresponding to \eqn{hgeq} is of the form \eqn{eiginfi}
and should correspond to the spectrum \eqn{hlspq} of \eqn{su2cft}.
The two spectra however differ, the reason being
the extra $\mathbb{R}$ factor in (4.46), as well as the fact that
the corresponding backgrounds are only locally equivalent
and differ in their global structure. To compare them,
we must first add an extra term $\delta^2/k_2$ to \eqn{hlspq},
corresponding to the momentum contribution in the extra
$\lambda$-direction in \eqn{hgeq} that does not appear in
\eqn{su2cft}. Noticing, further, that the periodic coordinate
in the change of variables \eqn{jdperi} is $\varphi + \lambda$,
we must also make the replacement $\delta \to \delta - m$
in this term. Upon doing that, we indeed obtain (4.41)
(with $m$ playing the role of $n$).

\no
Focusing around $\a_0=0$ in the high spin limit is not the only possibility.
Consider, instead, focusing around
$\a_0 =1$ and $\g =0$, by performing first the coordinate transformation
\be
\a_0^2 = 1 - r^2 \left[(x_1+\psi )^2 + x_3^2 \right]\ ,
\qq \g = r (x_1+\psi) \cos\psi\ , \qq \b_0 = \sin \psi\ ,
\ee
followed by the limit $r\to 0$. Then the new variables $x_1$ and $x_3$ that we will
use, instead of $\a_0$ and $\g$, become uncompacified.
In this limit we obtain for the metric and dilaton the expressions
\be
ds^2 =k_2\left(d\psi^2 + {\cos^2 \psi\ov x_3^2}d x_1^2
+ {\left( x_3 dx_3 + (\sin\psi\cos\psi +  x_1+\psi) dx_1\right)^2\ov x_3^2\cos^2\psi}\right)
\label{non1}
\ee
and
\be
e^{-2 \Phi} = x_3^2 \cos^2 \psi\ .
\label{non2}
\ee
It is easy to check that in this limit the state \eqn{psioo} becomes just a constant and trivially solves the
scalar wave equation \eqn{lappp} corresponding to the above background, with zero eigenvalue.
Similarly, the state \eqn{expp2}, which is insensitive to the high spin
limit \eqn{larspin2}, solves the same equation with eigenvalue $j(j+1)/k_2$.

\no
The above backgrounds, that is \eqn{kli1}, \eqn{kli2} and \eqn{non1}, \eqn{non2},
can be considered as the effective backgrounds describing the high spin sector of
the original CFT coset model.

\no
According to a general result derived in \cite{gwzwsfe} the above limit background \eqn{non1}, \eqn{non2}
should be the non-abelian dual of the $SU(2)_{k_2}$ WZW model with respect to the vectorial action
of $SU(2)$. Indeed, after a slight renaming of variables, it becomes identical to eqs. (6.10)
and (6.11) of \cite{GiRo}.
This suggests that non-abelian duals of WZW models (and possibly in other cases)
describe in fact consistent sectors of representations with very high values for the Casimir operators.
This is analogous to the plane-wave geometry \cite{Papado}
 describing the high mass and spin dimension sector of $\cN=4$
SYM within the AdS/CFT correspondence \cite{BMN}. The crucial step is to explicitly demonstrate that
the general state \eqn{pjhr1} has a well defined large spin limit that simultaneously solves the scalar wave
equation corresponding to the above limit background.
We plan to present work along these lines in the near future.% \cite{PolSfeapear}.

%\section{Extensions to $SU(N)_{k_1} \times SU(N)_{k_2} / SU(N)_{k_1+k_2}$ (\textbf{TO BE DONE ???)}}

\section{Concluding remarks and future directions}

We have presented a group theoretical method allowing for the explicit solution of field equations in
string backgrounds based on coset CFTs and the associated gauged WZW models. This is
an impossible task with the traditional methods of solving differential equations.
We have presented a general
formula for the scalar equation which we also tested explicitly for the non-trivial case of the
compact coset $SU(2)\times SU(2)/SU(2)$.

\no
Our work opens several possibilities concerning
physical applications of coset CFTs. In particular, one of the original motivations of such models
was to describe spacetimes with one-time coordinate which requires the use of
non-compact groups. In that respect, consider the class of $d$-dimensional
CFT coset models $SO(d-2,2)/SO(d-1,1)$ \cite{BaNe}, for which the background fields for the low dimensional cases
$d=3,4$ have been worked out explicitly \cite{BaSfe1,BaSfe2}. They can be given a spacetime
interpretation as anisotropic cosmological models. It will be very interesting to study field
propagation in these geometries and extract physical information. This is possible
using the methods we have developed, but in practice it requires a knowledge of the representation
theory of non-compact groups, especially in relation to
the specific physical application, which is a quite involved subject on its own.

\no
In \cite{PetroSfe} it was shown that there exists a hierarchy of nested CFT cosets which
schematically means that the smaller in dimensionality model reside at the boundary of the immediately
higher one and in fact it generates it by marginal deformations. It will be interesting to
shed further light on this relation in view of our present work.

\no
We conclude by pointing out that the scalar equation for
conformal cosets of the form $SU(N) \times SU(N) / SU(N)$, similar
to the ones we considered in sections 3 and 4, is closely related to the singlet sector of unitary two-matrix models. The singlet sector
of the unitary one-matrix model based on $SU(N)/SU(N)$ is known to
lead to free fermions, while non-singlet sectors correspond to
generalizations of the spin-Calogero-Sutherland integrable model
(see \cite{caloreview} for a review).
Specific sectors of a unitary many-matrix model are also known to
give rise to further generalizations of the above integrable model
with anisotropic spin couplings \cite{manymatrix}, while $G/G$
WZW models are related to its relativistic version \cite{nekra}.
Implementation of the methods in the present paper may lead to further
connections between WZW models and generalized integrable models.

\vskip .4 in
\centerline{ \bf Acknowledgments}

\no
K.S. would like to thank S.D. Avramis for participating in preliminary stages of this research.
In addition, he would like to thank the City College of New York for hospitality and financial support
during a visit in which part of this work was done.

%%%%%%%%%%%%%%%%%%%%%%%%%%%%%%%%%%%%%%%%%%%%%%%%%%%%%%%%%%%%%%%%%%

\end{document}